\documentclass[12pt]{article}
\usepackage{a4wide,epsfig,epsf}

\catcode`@=11
\newcount\@tempcntc
\def\@citex[#1]#2{\if@filesw\immediate\write\@auxout{\string\citation{#2}}\fi
  \@tempcnta\z@\@tempcntb\m@ne\def\@citea{}\@cite{\@for\@citeb:=#2\do
    {\@ifundefined
       {b@\@citeb}{\@citeo\@tempcntb\m@ne\@citea\def\@citea{,}{\bf
?}\@warning
       {Citation `\@citeb' on page \thepage \space undefined}}%
    {\setbox\z@\hbox{\global\@tempcntc0\csname b@\@citeb\endcsname\relax}%
     \ifnum\@tempcntc=\z@ \@citeo\@tempcntb\m@ne
       \@citea\def\@citea{,}\hbox{\csname b@\@citeb\endcsname}%
     \else
      \advance\@tempcntb\@ne
      \ifnum\@tempcntb=\@tempcntc
      \else\advance\@tempcntb\m@ne\@citeo
      \@tempcnta\@tempcntc\@tempcntb\@tempcntc\fi\fi}}\@citeo}{#1}}
\def\@citeo{\ifnum\@tempcnta>\@tempcntb\else\@citea\def\@citea{,}%
  \ifnum\@tempcnta=\@tempcntb\the\@tempcnta\else
   {\advance\@tempcnta\@ne\ifnum\@tempcnta=\@tempcntb \else
\def\@citea{--}\fi
    \advance\@tempcnta\m@ne\the\@tempcnta\@citea\the\@tempcntb}\fi\fi}
\catcode`@=12

\begin{document}
\title{\vskip-3cm{\baselineskip14pt
\centerline{\normalsize DESY 02-168\hfill ISSN 0418-9833}
\centerline{\normalsize hep-ph/0306080\hfill}
\centerline{\normalsize June 2003\hfill}}
\vskip1.5cm
Charmonium production in polarized high-energy collisions}
\author{{\sc M. Klasen, B. A. Kniehl, L. N. Mihaila, M. Steinhauser}\\
{\normalsize II. Institut f\"ur Theoretische Physik, Universit\"at
Hamburg,}\\
{\normalsize Luruper Chaussee 149, 22761 Hamburg, Germany}}

\date{}

\maketitle

\thispagestyle{empty}

\begin{abstract}
We investigate the inclusive production of prompt $J/\psi$ mesons in polarized
hadron-hadron, photon-hadron, and photon-photon collisions in the
factorization formalism of nonrelativistic quantum chromodynamics (NRQCD)
providing all contributing partonic cross sections in analytic form.
In the case of photoproduction, we also include the resolved-photon
contributions.
We present numerical results appropriate for BNL RHIC-Spin, the approved SLAC
fixed-target experiment E161, and the $e^+e^-$ and $\gamma\gamma$ modes of
TESLA.
Specifically, we assess the feasibility to access the spin-dependent parton
distributions in the polarized proton and photon.
We also point out that preliminary data on $J/\psi$ inclusive production taken
by the PHENIX Collaboration in unpolarized proton-proton collisions at RHIC
tends to favor the NRQCD factorization hypothesis, while it significantly
overshoots the theoretical prediction of the color-singlet model at large
values of transverse momentum.

\medskip

\end{abstract}

\newpage

\section{Introduction}

Since its discovery in 1974, the $J/\psi$ meson has provided a useful
laboratory for quantitative tests of quantum chromodynamics (QCD) and, in
particular, of the interplay of perturbative and nonperturbative phenomena.
The factorization formalism of nonrelativistic QCD (NRQCD) \cite{bbl} provides
a rigorous theoretical framework for the description of heavy-quarkonium
production and decay.
This formalism implies a separation of short-distance coefficients, which can 
be calculated perturbatively as expansions in the strong-coupling constant
$\alpha_s$, from long-distance matrix elements (MEs), which must be extracted
from experiment.
The relative importance of the latter can be estimated by means of velocity
scaling rules, {\it i.e.}, the MEs are predicted to scale with a definite
power of the heavy-quark ($Q$) velocity $v$ in the limit $v\ll1$.
In this way, the theoretical predictions are organized as double expansions in
$\alpha_s$ and $v$.
A crucial feature of this formalism is that it takes into account the complete
structure of the $Q\overline{Q}$ Fock space, which is spanned by the states
$n={}^{2S+1}L_J^{(c)}$ with definite spin $S$, orbital angular momentum $L$,
total angular momentum $J$, and color multiplicity $c=1,8$.
In particular, this formalism predicts the existence of color-octet (CO)
processes in nature.
This means that $Q\overline{Q}$ pairs are produced at short distances in
CO states and subsequently evolve into physical, color-singlet (CS) quarkonia
by the nonperturbative emission of soft gluons.
In the limit $v\to0$, the traditional CS model (CSM) \cite{ber,gas} is
recovered.
The greatest triumph of this formalism was that it was able to correctly 
describe \cite{bra} the cross section of inclusive charmonium
hadroproduction measured in $p\overline{p}$ collisions at the Fermilab
Tevatron \cite{abe}, which had turned out to be more than one order of
magnitude in excess of the theoretical prediction based on the CSM.

Apart from this phenomenological drawback, the CSM also suffers from severe
conceptual problems indicating that it is incomplete.
These include the presence of logarithmic infrared divergences in the
${\cal O}(\alpha_s)$ corrections to $P$-wave decays to light hadrons and in
the relativistic corrections to $S$-wave annihilation \cite{bar}, and the lack
of a general argument for its validity in higher orders of perturbation
theory.
While the $k_T$-factorization \cite{sri} and hard-comover-scattering
\cite{hoy} approaches manage to bring the CSM prediction much closer to the
Tevatron data, they do not cure the conceptual defects of the CSM.
The color evaporation model \cite{cem}, which is intuitive and useful for
qualitative studies, also leads to a significantly better description of the
Tevatron data, but it is not meant to represent a rigorous framework for
perturbation theory.
In this sense, a coequal alternative to the NRQCD factorization formalism is 
presently not available.

In order to convincingly establish the phenomenological significance of the
CO processes, it is indispensable to identify them in other kinds of
high-energy experiments as well.
Studies of charmonium production in $ep$ photoproduction, $ep$ and $\nu N$
deep-inelastic scattering (DIS), $e^+e^-$ annihilation, $\gamma\gamma$
collisions, and $b$-hadron decays may be found in the literature; see
Ref.~\cite{fle} and references cited therein.
Furthermore, the polarization of charmonium, which also provides a sensitive
probe of CO processes, was investigated \cite{ben,bkl}.
Until very recently, none of these studies was able to prove or disprove the
NRQCD factorization hypothesis.
However, H1 data of $ep\to eJ/\psi+X$ in DIS at HERA \cite{h1} and DELPHI data
of $\gamma\gamma\to J/\psi+X$ at LEP2 \cite{delphi} provide first independent
evidence for it \cite{dis,gg}.

Before the advent of the NRQCD factorization formalism, heavy-quarkonium
production in experiments with polarized proton or photon beams was believed
to provide reliable information on the spin-dependent gluon distribution
functions of the polarized proton or photon.
At present, however, we are faced with the potential problem that NRQCD
predictions at lowest order (LO) come with a considerable normalization
uncertainty.
This is due to the additional expansion in $v$, whose convergence property has
yet to be tested, and the introduction of CO MEs as additional input
parameters.
In fact, the relevant CO MEs have only been determined through LO fits to
experimental data and thus come with appreciable theoretical uncertainties.
Thus, it must be clarified if heavy-quarkonium production with polarized
proton or photon beams remains to be a useful probe of the polarized gluon
distribution functions.
It is the purpose of this paper to answer this question.
Specifically, we consider inclusive $J/\psi$ production in polarized $pp$,
$\gamma p$, and $\gamma\gamma$ collisions, appropriate for RHIC-Spin, 
experiment E161, and the DESY TeV-Energy Superconducting Linear Accelerator
(TESLA) operated in the $e^+e^-$ and $\gamma\gamma$ modes, respectively.
In the RHIC-Spin mode of the Relativistic Heavy Ion Collider (RHIC) at
Brookhaven National Laboratory (BNL), proton beams with strong longitudinal
polarization, of approximately 70\%, collide with center-of-mass (c.m.) energy
up to $\sqrt S=500$~GeV and luminosity
${\cal L}=2\times10^{32}$~cm$^{-2}$s$^{-1}$ \cite{bun}.
In the approved experiment E161 at the Stanford Linear Accelerator Center
(SLAC), circularly polarized photons with energies between 35 and 48~GeV will
collide on a fixed target made of longitudinally polarized deuterium 
\cite{bos}.
At TESLA, which is being designed and planned at the German Electron
Synchrotron (DESY) in Hamburg, longitudinal electron and positron
polarizations of up to 80\% and 60\%, respectively, can be achieved.
In the $e^+e^-$ mode \cite{tesla}, photons are unavoidably generated by hard
initial-state bremsstrahlung and by beamstrahlung, the synchrotron radiation
emitted by one of the colliding bunches in the field of the opposite bunch.
In both cases, polarization is transferred from the electrons and positrons to
the photons in a calculable way.
TESLA can also be converted into a photon collider via Compton back-scattering
of high-energetic laser light off the electron beams \cite{compton}.
With 100\% circular laser polarization and up to 80\% longitudinal electron
polarization, the scattered photons can be arranged to have a strong
circular polarization, especially at the upper and lower ends of their energy
spectrum.

The photons can interact either directly with the quarks participating in the
hard-scattering process (direct photoproduction) or via their quark and gluon
content (resolved photoproduction).
Thus, the process $\gamma\gamma\to J/\psi+X$, where $X$ collectively denotes
all unobserved particles produced along with the $J/\psi$ meson, receives
contributions from direct, single-resolved, and double-resolved channels.
All three contributions are formally of the same order in the perturbative
expansion and must be included.
This may be understood by observing that the parton distribution functions
(PDFs) of the photon have a leading behavior proportional to
$\alpha\ln\left(\mu_f^2/\Lambda_{\rm QCD}^2\right)\propto\alpha/\alpha_s$,
where $\alpha$ is Sommerfeld's fine-structure constant, $\mu_f$ is the
factorization scale, and $\Lambda_{\rm QCD}$ is the asymptotic scale parameter
of QCD.
Similarly, the reaction $\gamma N\to J/\psi+X$, where $N$ is a nucleon,
proceeds via direct and resolved channels, which must all be taken into 
account.

The $J/\psi$ mesons can be produced directly; or via radiative or hadronic
decays of heavier charmonia, such as $\chi_{cJ}$ and $\psi^\prime$ mesons; or
via weak decays of $b$ hadrons.
The respective decay branching fractions are
$B(\chi_{c0}\to J/\psi+\gamma)=(1.02\pm0.17)\%$,
$B(\chi_{c1}\to J/\psi+\gamma)=(31.6\pm3.2)\%$,
$B(\chi_{c2}\to J/\psi+\gamma)=(18.7\pm2.0)\%$,
$B(\psi^\prime\to J/\psi+X)=(55.7\pm2.6)\%$, and
$B(B\to J/\psi+X)=(1.15\pm0.06)\%$ \cite{pdg}.
At RHIC-Spin and TESLA, the $b$ hadrons can be detected by looking for
displaced decay vertices with dedicated vertex detectors.
At RHIC-Spin, only about 1\% of all $J/\psi$ mesons are expected to originate
from $b$-hadron decays \cite{sat}.
In experiment E161, the c.m.\ energy is not sufficient to produce
$b\overline{b}$ pairs.
In our analysis, we will not consider $J/\psi$ mesons from $b$-hadron decays.
The cross sections of the four residual indirect production channels may be
approximated by multiplying the direct-production cross sections of the
respective intermediate charmonia with their decay branching fractions to
$J/\psi$ mesons.
It has become customary to collectively denote the $J/\psi$ mesons produced
directly or via the feed-down from heavier charmonia as prompt.

This paper is organized as follows.
In Sec.~\ref{sec:ana}, we list our formulas for the contributing partonic
cross sections and compare them with the available literature results.
Lengthy expressions are relegated to Appendix.
In Sec.~\ref{sec:num}, we present our numerical results and discuss their
phenomenological implications for the three experiments under consideration.
Our conclusions are summarized in Sec.~\ref{sec:con}.

\section{\label{sec:ana}Analytical results}

Let us consider the generic inclusive process $AB\to H+X$, where $A$ and $B$
are the beam particles (hadrons or photons) with definite helicities $\xi_A$
and $\xi_B$, respectively, and $H$ is the charmonium state, and let
$\sigma^{\xi_A,\xi_B}$ denote its cross section.
Then, the double longitudinal-spin asymmetry is defined as
\begin{equation}
{\cal A}_{LL}=\frac{\Delta\sigma}{\sigma},
\end{equation}
where
\begin{eqnarray}
\sigma&=&\frac{1}{4}\sum_{\xi_A,\xi_B=\pm1}\sigma^{\xi_A,\xi_B},
\nonumber\\
\Delta\sigma&=&\frac{1}{4}\sum_{\xi_A,\xi_B=\pm1}(-1)^{\xi_A\xi_B}\,
\sigma^{\xi_A,\xi_B}
\end{eqnarray}
are the unpolarized and polarized cross sections, respectively.
An alternative definition of ${\cal A}_{LL}$, which allows one to study the
dependences on the transverse momentum $p_T$ and rapidity $y$ of $H$, uses the
differential cross sections $d\Delta\sigma$ and $d\sigma$.

Invoking the factorization theorems of the QCD parton model and NRQCD, the
differential polarized cross section can be written as
\begin{eqnarray}
d\Delta\sigma(AB\to H+X)&=&\sum_{a,b,d}
\int dx_a\Delta f_{a/A}(x_a,\mu_f)\int dx_b\Delta f_{b/B}(x_b,\mu_f)
\sum_n\langle{\cal O}^H[n]\rangle
\nonumber\\
&&{}\times d\Delta\sigma(ab\to c\overline{c}[n]d),
\label{eq:dds}
\end{eqnarray}
where it is summed over $a,b=\gamma,g,q,\overline{q}$ and 
$d=g,q,\overline{q}$,
$x_a$ is the fraction of longitudinal momentum that $a$ receives from $A$,
$\mu_f$ is the factorization scale, $\langle{\cal O}^H[n]\rangle$ are the MEs
of the $H$ meson, and
\begin{eqnarray}
\Delta f_{a/A}(x_a,\mu_f)&=&f_{a/A}^+(x_a,\mu_f)-f_{a/A}^-(x_a,\mu_f),
\nonumber\\
d\Delta\sigma(ab\to c\overline{c}[n]d)&=&
\frac{1}{2}\left[d\sigma^+(ab\to c\overline{c}[n]d)
-d\sigma^-(ab\to c\overline{c}[n]d)\right].
\end{eqnarray}
Here, $f_{a/A}^\pm(x_a,\mu_f)$ are the PDFs of $A$ for the case that the spin
of $a$ is (anti)parallel to that of $A$ ($\xi_A\xi_a=\pm1$), and
$d\sigma^\pm(ab\to c\overline{c}[n]d)$ refers to the case of
$\xi_a\xi_b=\pm1$.
In fact, due to parity conservation in QCD,
$d\sigma^\pm(ab\to c\overline{c}[n]d)$ only depends on the product
$\xi_a\xi_b$, as can be seen from the Appendix.
With the definition
$\Delta f_{\gamma/\gamma}(x_\gamma,\mu_f)=\delta(1-x_\gamma)$,
Eq.~(\ref{eq:dds}) accommodates the direct, single-resolved, and
double-resolved channels.
In the case of TESLA, additional convolutions with appropriate photon flux
functions $\Delta f_{\gamma/e}(x)=f_{\gamma/e}^+(x)-f_{\gamma/e}^-(x)$, where 
the superscript $\pm$ refers to the case of $\xi_e\xi_\gamma=\pm1$, have to be
implemented in Eq.~(\ref{eq:dds}), invoking the Weizs\"acker-Williams
approximation (WWA) \cite{wwa} for electromagnetic bremsstrahlung and its
analogues for beamstrahlung and Compton back-scattering.
The corresponding formulas for the unpolarized case are obtained by omitting
everywhere in Eq.~(\ref{eq:dds}) the $\Delta$ symbols.
The kinematical relations needed to calculate $p_T$ and $y$ distributions from
Eq.~(\ref{eq:dds}) may be found, {\it e.g.}, in Ref.~\cite{kpz}.

We work in the fixed-flavor-number scheme, {\it i.e.}, we have $n_f=3$ active
quark flavors $q=u,d,s$ in the proton and resolved photon.
As required by parton-model kinematics, we treat the quarks $q$ as massless.
The charm quark $c$ and antiquark $\overline{c}$ only appear in the final 
state.
The $c\overline{c}$ Fock states contributing at LO in $v$ are
$n={}^1\!S_0^{(1)},{}^1\!S_0^{(8)},{}^3\!S_1^{(8)},{}^1\!P_1^{(8)}$ if
$H=\eta_c$;
$n={}^3\!S_1^{(1)},{}^1\!S_0^{(8)},{}^3\!S_1^{(8)},{}^3\!P_J^{(8)}$ if
$H=J/\psi,\psi^\prime,\psi(3S),\ldots$;
$n={}^1\!P_1^{(1)},{}^1\!S_0^{(8)}$ if $H=h_c$; and
$n={}^3\!P_J^{(1)},{}^3\!S_1^{(8)}$ if $H=\chi_{cJ}$, where $J=0,1,2$.
Their MEs satisfy the multiplicity relations
\begin{eqnarray}
\left\langle{\cal O}^{\psi(nS)}\left[{}^3\!P_J^{(8)}\right]\right\rangle
&=&(2J+1)
\left\langle{\cal O}^{\psi(nS)}\left[{}^3\!P_0^{(8)}\right]\right\rangle,
\nonumber\\
\left\langle{\cal O}^{\chi_{cJ}}\left[{}^3\!P_J^{(1)}\right]\right\rangle
&=&(2J+1)
\left\langle{\cal O}^{\chi_{c0}}\left[{}^3\!P_0^{(1)}\right]\right\rangle,
\nonumber\\
\left\langle{\cal O}^{\chi_{cJ}}\left[{}^3\!S_1^{(8)}\right]\right\rangle
&=&(2J+1)
\left\langle{\cal O}^{\chi_{c0}}\left[{}^3\!S_1^{(8)}\right]\right\rangle,
\label{eq:mul}
\end{eqnarray}
which follow to LO in $v$ from heavy-quark spin symmetry.
In our numerical analysis, we only include the $J/\psi$, $\chi_{cJ}$, and
$\psi^\prime$ mesons.
For completeness and future use, we also list formulas for all the other
experimentally established charmonia.
Our results readily carry over to all known bottomonia as well.
In order for the $J/\psi$ meson to have finite transverse momentum $p_T$, we
allow for an additional parton (quark or gluon) in the final state.
We do not include the $J/\psi+\gamma$ final state because prompt photons can
be identified experimentally, so that such events can be eliminated from the
data sample.

We are thus led to consider the following partonic subprocesses
\begin{eqnarray}
\gamma\gamma&\to&c\overline{c}[\varsigma^{(8)}]g,
\label{eq:ppog}\\
\gamma g&\to&c\overline{c}[\varsigma^{(1)}]g,
\label{eq:pgsg}\\
\gamma g&\to&c\overline{c}[\varsigma^{(8)}]g,
\label{eq:pgog}\\
\gamma q&\to&c\overline{c}[\varsigma^{(8)}]q,
\label{eq:pqoq}\\
gg&\to&c\overline{c}[\varsigma^{(1)}]g,
\label{eq:ggsg}\\
gg&\to&c\overline{c}[\varsigma^{(8)}]g,
\label{eq:ggog}\\
gq&\to&c\overline{c}[\varsigma^{(1)}]q,
\label{eq:gqsq}\\
gq&\to&c\overline{c}[\varsigma^{(8)}]q,
\label{eq:gqoq}\\
q\overline{q}&\to&c\overline{c}[\varsigma^{(1)}]g,
\label{eq:qqsg}\\
q\overline{q}&\to&c\overline{c}[\varsigma^{(8)}]g,
\label{eq:qqog}
\end{eqnarray}
where $\varsigma={}^1\!S_0,{}^3\!S_1,{}^1\!P_1,{}^3\!P_J$ with $J=0,1,2$.
The processes $\gamma\gamma\to c\overline{c}[\varsigma^{(1)}]g$ are forbidden
by color conservation.
Similarly, the processes $\gamma q\to c\overline{c}[\varsigma^{(1)}]q$ are
prohibited because the $c$-quark line is connected with the $q$-quark line
by one gluon, which transmits color to the $c\overline{c}$ pair.
By angular-momentum conservation, processes (\ref{eq:ppog}) and
(\ref{eq:pgsg}) with $\varsigma={}^1\!S_0,{}^3\!P_J$, process (\ref{eq:pqoq})
with $\varsigma={}^1\!P_1$, and processes (\ref{eq:gqsq}) and (\ref{eq:qqsg})
with $\varsigma={}^3\!S_1,{}^1\!P_1$ have zero cross sections.
We calculated the differential cross sections $d\sigma/dt$ of all remaining
partonic subprocesses for arbitrary helicities $\xi_a,\xi_b=\pm1$ of the
incoming particles $a,b=\gamma,g,q,\overline{q}$.
Our formulas are listed in the Appendix.

In the literature, inclusive charmonium photoproduction with polarized beams
was studied both in the CSM \cite{duk,mor} and in NRQCD \cite{yua,jap,sud}.
In the case of hadroproduction, CSM and NRQCD studies may be found in
Refs.~\cite{gas,mor,rob,don} and Ref.~\cite{tka}, respectively.
We are not aware of any previous analyses for collisions of polarized photon
beams.
Some of these papers contain analytic results \cite{gas,duk,mor,yua,jap,rob},
which we can compare with ours.
This is necessary because the literature results are in some cases mutually
conflicting.
As for direct $J/\psi$ photoproduction, our result for the CS process
(\ref{eq:pgsg}) with $\varsigma={}^3\!S_1$ agrees with Refs.~\cite{mor,yua},
while it disagrees with Ref.~\cite{duk}; our results for the CO processes
(\ref{eq:pgog}) and (\ref{eq:pqoq}) with
$\varsigma={}^1\!S_0,{}^3\!S_1,{}^3\!P_J$ agree with
Ref.~\cite{yua}\footnote{%
We do not agree with Eqs.~(A6) and (A7) of Ref.~\cite{yua} and find an overall
minus sign relative to Eq.~(A11).
However, we agree with Eqs.~(A12) and (A13), which refer to the sum over the
three ${}^3\!P_J^{(8)}$ states.}
and partly with Ref.~\cite{jap}.\footnote{%
We agree with Eq.~(A3) of Ref.~\cite{jap} if we divide the right-hand side of
this equation by $s^2$.
Equations~(A4)--(A6) are corrupted, contain undefined symbols, and are thus
unsuitable for comparisons.}
As for charmonium hadroproduction in the CSM, our results for processes
(\ref{eq:ggsg}) agree with Refs.~\cite{gas,rob} and those for processes
(\ref{eq:gqsq}) and (\ref{eq:qqsg}) with $\varsigma={}^3\!P_J$ agree with
Ref.~\cite{rob}.
To our knowledge, our residual formulas are not available in the literature.

\section{\label{sec:num}Numerical results}

We are now in a position to present our numerical results.
We first describe our theoretical input and the kinematic conditions.
We use $m_c=(1.5\pm0.1)$~GeV, $\alpha=1/137.036$, and the LO formula for
$\alpha_s^{(n_f)}(\mu)$ \cite{pdg} with $n_f=3$ active quark flavors and
asymptotic scale parameter $\Lambda^{(3)}=204$~MeV.
Here, $m_c$ denotes the charm-quark mass.

As for the polarized proton PDFs, we use the LO sets by Gl\"uck, Reya,
Stratmann, and Vogelsang (GRStV) \cite{grstv} and those by Gehrmann and
Stirling (GS) \cite{gs}.
Their reference sets of unpolarized proton PDFs are both by Gl\"uck, Reya, and 
Vogt, namely the LO sets of Refs.~\cite{grv98} (GRV98) and \cite{grv95}
(GRV95), respectively.
There are two GRStV sets, one of which corresponds to the common
{\it standard} (STD) scenario of polarized PDFs with a flavor-symmetric
light-sea (antiquark) distribution $\Delta f_{\overline{q}/p}$, while the
other one refers to a completely $SU(3)_f$-broken {\it valence} (VAL) scenario
with totally flavor-asymmetric light-sea distributions
($\Delta f_{\overline{u}/p}\ne\Delta f_{\overline{d}/p}
\ne\Delta f_{\overline{s}/p}$).
The latter are modelled with the help of a Pauli-blocking ansatz at the low
radiative or dynamical input scale $\mu_{\rm LO}^2=0.26$~GeV$^2$, which 
complies with predictions of the chiral quark-soliton model and expectations
based on the statistical parton model as well as with the corresponding 
flavor-broken unpolarized sea ($f_{\overline{d}/p}>f_{\overline{u}/p}$).
There are three GS sets (A,B,C), corresponding to three different, equally
possible scenarios for the polarized gluon distribution, which is only loosely
constrained by current experimental data.
Leaving aside nuclear corrections and appealing to strong-isospin symmetry,
the effective nucleon ($N$) PDFs may be approximated by
\begin{equation}
f_{a/N}(x,\mu_f)
=\frac{1}{A}\left[Zf_{a/p}(x,\mu_f)+(A-Z)f_{b/p}(x,\mu_f)\right],
\end{equation}
and similarly for $\Delta f_{a/N}(x,\mu_f)$, where $Z$ and $A$ are the
respective numbers of protons and nucleons in the nucleus and
$(a,b)=(u,d),(\overline{u},\overline{d}),(d,u),(\overline{d},\overline{u}),
(s,s),(\overline{s},\overline{s}),(g,g)$.
In the case of deuterium $D$, we have $Z=1$ and $A=2$.
We present the cross sections per nucleon, rather than per nucleus.

Up-to-date LO sets of polarized photon PDFs were presented by Gl\"uck, Reya,
and Sieg (GRSi) \cite{grsi}.
Their unpolarized counterpart is the LO set by Gl\"uck, Reya, and Schienbein
(GRSc) \cite{grsc}.
There are two GRSi sets, which differ in the assumed boundary condition for
the polarized gluon distribution at the dynamical input scale and are 
characterized as maximally (MAX) or minimally (MIN) saturated.
Our default sets are taken to be GRStV-STD and GRSi-MAX as well as their
unpolarized counterparts GRV98 and GRSc, respectively.
The other sets are employed to assess the sensitivity of the considered
experiments to the details of the polarized proton and photon PDFs given the
theoretical uncertainties from other sources, especially those introduced by
NRQCD.

We choose the renormalization and factorization scales to be
$\mu_i=\xi_i\sqrt{4m_c^2+p_T^2}$, with $i=r,f$, respectively, and
independently vary the scale parameters $\xi_r$ and $\xi_f$ between 1/2 and 2
about the default value 1.
As for the $J/\psi$, $\chi_{cJ}$, and $\psi^\prime$ MEs, we adopt the set
determined in Ref.~\cite{bkl} using the LO proton PDFs by Martin, Roberts,
Stirling, and Thorne \cite{mrst}.
Specifically,
$\left\langle{\cal O}^{\psi(nS)}\left[{}^3\!S_1^{(1)}\right]\right\rangle$
with $n=1,2$ and
$\left\langle{\cal O}^{\chi_{c0}}\left[{}^3\!P_0^{(1)}\right]\right\rangle$
were extracted from the measured partial decay widths of $\psi(nS)\to l^+l^-$
and $\chi_{c2}\to\gamma\gamma$ \cite{pdg}, respectively, while
$\left\langle{\cal O}^{\psi(nS)}\left[{}^1\!S_0^{(8)}\right]\right\rangle$,
$\left\langle{\cal O}^{\psi(nS)}\left[{}^3\!S_1^{(8)}\right]\right\rangle$,
$\left\langle{\cal O}^{\psi(nS)}\left[{}^3\!P_0^{(8)}\right]\right\rangle$,
and
$\left\langle{\cal O}^{\chi_{c0}}\left[{}^3\!S_1^{(8)}\right]\right\rangle$
were fitted to the $p_T$ distributions of $\psi(nS)$ and $\chi_{cJ}$ inclusive
hadroproduction \cite{abe} and the cross-section ratio
$\sigma_{\chi_{c2}}/\sigma_{\chi_{c1}}$ \cite{aff} measured at the Tevatron.
The fit results for
$\left\langle{\cal O}^{\psi(nS)}\left[{}^1\!S_0^{(8)}\right]\right\rangle$ and
$\left\langle{\cal O}^{\psi(nS)}\left[{}^3\!P_0^{(8)}\right]\right\rangle$ are
strongly correlated, so that the linear combinations
\begin{equation}
M_r^{\psi(nS)}
=\left\langle{\cal O}^{\psi(nS)}\left[{}^1\!S_0^{(8)}\right]\right\rangle
+\frac{r}{m_c^2}
\left\langle{\cal O}^{\psi(nS)}\left[{}^3\!P_0^{(8)}\right]\right\rangle,
\label{eq:mr}
\end{equation}
with suitable values of $r$, are quoted.
Unfortunately, the hadronic cross sections under consideration here are
sensitive to different linear combinations of
$\left\langle{\cal O}^{\psi(nS)}\left[{}^1\!S_0^{(8)}\right]\right\rangle$ and
$\left\langle{\cal O}^{\psi(nS)}\left[{}^3\!P_0^{(8)}\right]\right\rangle$
than those appearing in Eq.~(\ref{eq:mr}).
In want of more specific information, we write
$\left\langle{\cal O}^{\psi(nS)}\left[{}^1\!S_0^{(8)}\right]\right\rangle
=\kappa_{\psi(nS)}M_r^{\psi(nS)}$
and
$\left\langle{\cal O}^{\psi(nS)}\left[{}^3\!P_0^{(8)}\right]\right\rangle
=(1-\kappa_{\psi(nS)})\left(m_c^2/r\right)M_r^{\psi(nS)}$ and independently
vary $\kappa_{J/\psi}$ and $\kappa_{\psi^\prime}$ between 0 and 1 around the
default value 1/2.
Our choice of $\Lambda^{(3)}$ coincides with the value employed in
Refs.~\cite{grstv,grv98,grsi,grsc,mrst}, and it is very close to the value
232~MeV of Refs.~\cite{gs,grv95}.

In order to estimate the theoretical uncertainties in our predictions, we
vary the unphysical parameters $\xi_r$, $\xi_f$, $\kappa_{J/\psi}$, and
$\kappa_{\psi^\prime}$ as indicated above and take into account the
experimental errors on $m_c$, the decay branching fractions, and the MEs.
We then combine the individual shifts in quadrature, allowing for the upper
and lower half-errors to be different.

We now discuss the photon flux functions that enter our predictions for
photon-photon scattering at TESLA.
The energy spectrum of the bremsstrahlung photons is well described in the 
WWA.
The formulas for the unpolarized and polarized cases, may be found in Eq.~(20)
of Ref.~\cite{fri} and Eq.~(14) of Ref.~\cite{def}, respectively, where the
maximum photon virtuality $Q^2$ is given in Eq.~(26) of Ref.~\cite{fri}.
We assume that the scattered electrons and positrons will be antitagged, as
was usually the case at LEP2, and take the maximum scattering angle to be
$\theta_{\rm max}=25$~mrad \cite{theta}.
The energy spectrum of the beamstrahlung photons is approximately described by
Eq.~(2.14) of Ref.~\cite{che} in the unpolarized case and by Eq.~(7.10) of
Ref.~\cite{kla} (see also Ref.~\cite{sch}) in the polarized one.
It is controlled by the effective beamstrahlung parameter $\Upsilon$, which is
given by Eq.~(2.10) of Ref.~\cite{che}.
Inserting the relevant TESLA parameters for the $\sqrt S=500$~GeV baseline
design specified in Table~1.3.1 of Ref.~\cite{tesla} in that formula, we
obtain $\Upsilon=0.053$.
In the case of the $e^+e^-$ mode of TESLA, we coherently superimpose the WWA
and beamstrahlung spectra.
Finally, in the case of the $\gamma\gamma$ mode of TESLA, the energy spectrum
of the back-scattered laser photons is given by Eq.~(6a) of Ref.~\cite{gin} in
the unpolarized case and by Eq.~(4) of Ref.~\cite{kot} in the polarized one.
It depends on the parameter $\kappa=s_{e\gamma}/m_e^2-1$, where
$\sqrt{s_{e\gamma}}$ is the c.m.\ energy of the charged lepton and the laser
photon, and it extends up to $x_{\rm max}=\kappa/(\kappa+1)$, where $x$ is the
energy of the back-scattered photons in units of $\sqrt S/2$.
The optimal value of $\kappa$ is $\kappa=2\left(1+\sqrt2\right)\approx4.83$
\cite{gin}, which we adopt; for larger values of $\kappa$, $e^+e^-$ pairs
would be created in the collisions of laser and back-scattered photons.

\begin{figure}[ht]
\begin{center}
\epsfig{figure=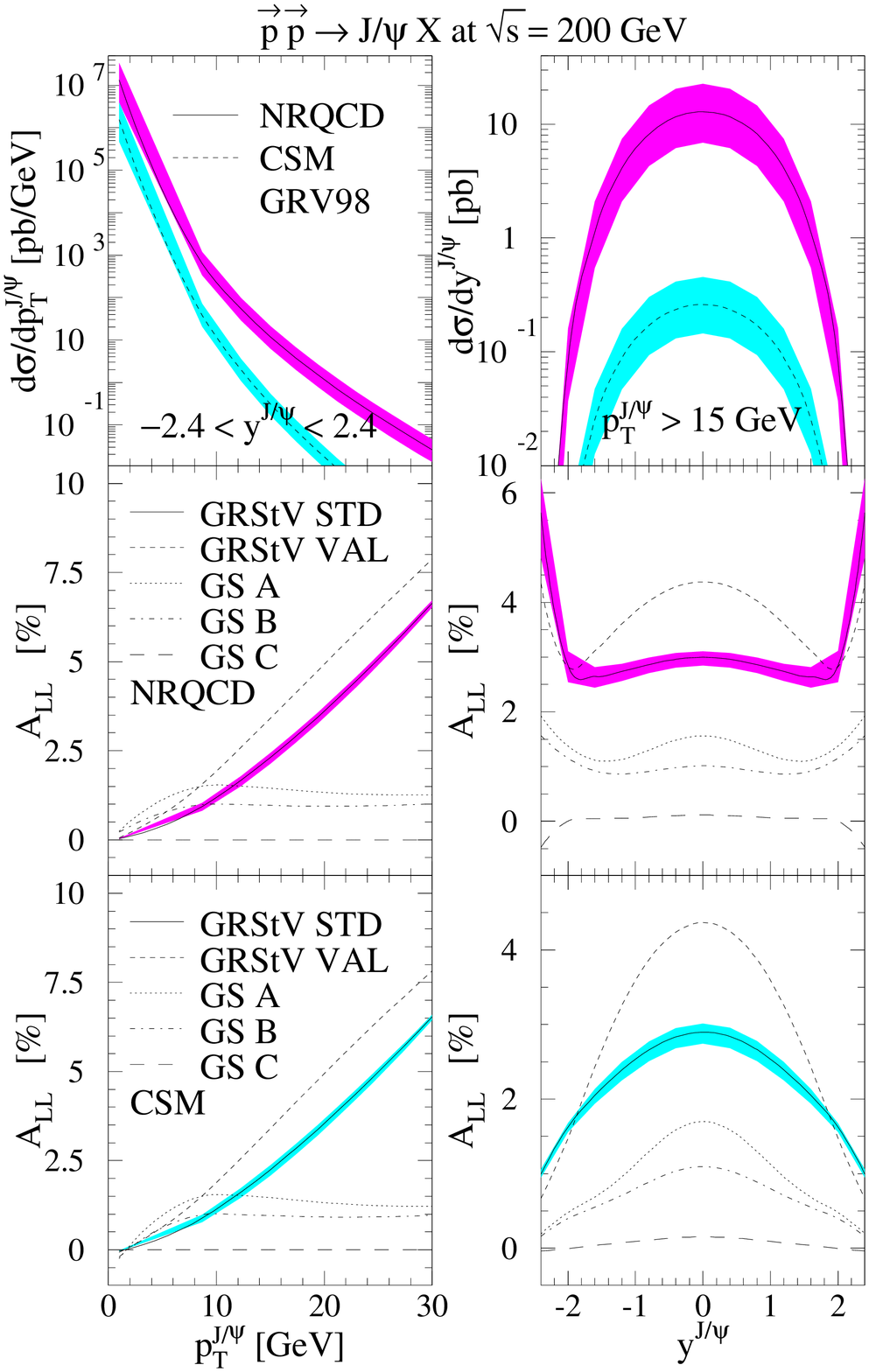,height=18cm}
\caption{The unpolarized cross sections $d\sigma/dp_T$ and $d\sigma/dy$
(first panel) and the double longitudinal-spin asymmetry ${\cal A}_{LL}$
(second and third panels) of $pp\to J/\psi+X$ at RHIC-Spin, with
$\sqrt S=200$~GeV, are studied as functions of $p_T$ (left column) and $y$
(right column) in NRQCD and the CSM.
The shaded bands indicate the theoretical uncertainties in the NRQCD and CSM
predictions evaluated with the default PDFs.
In the case of ${\cal A}_{LL}$, these uncertainties are compared with the
spread due to a variation of the polarized PDFs.
\label{fig:rhic2}}
\end{center}
\end{figure}

\begin{figure}[ht]
\begin{center}
\epsfig{figure=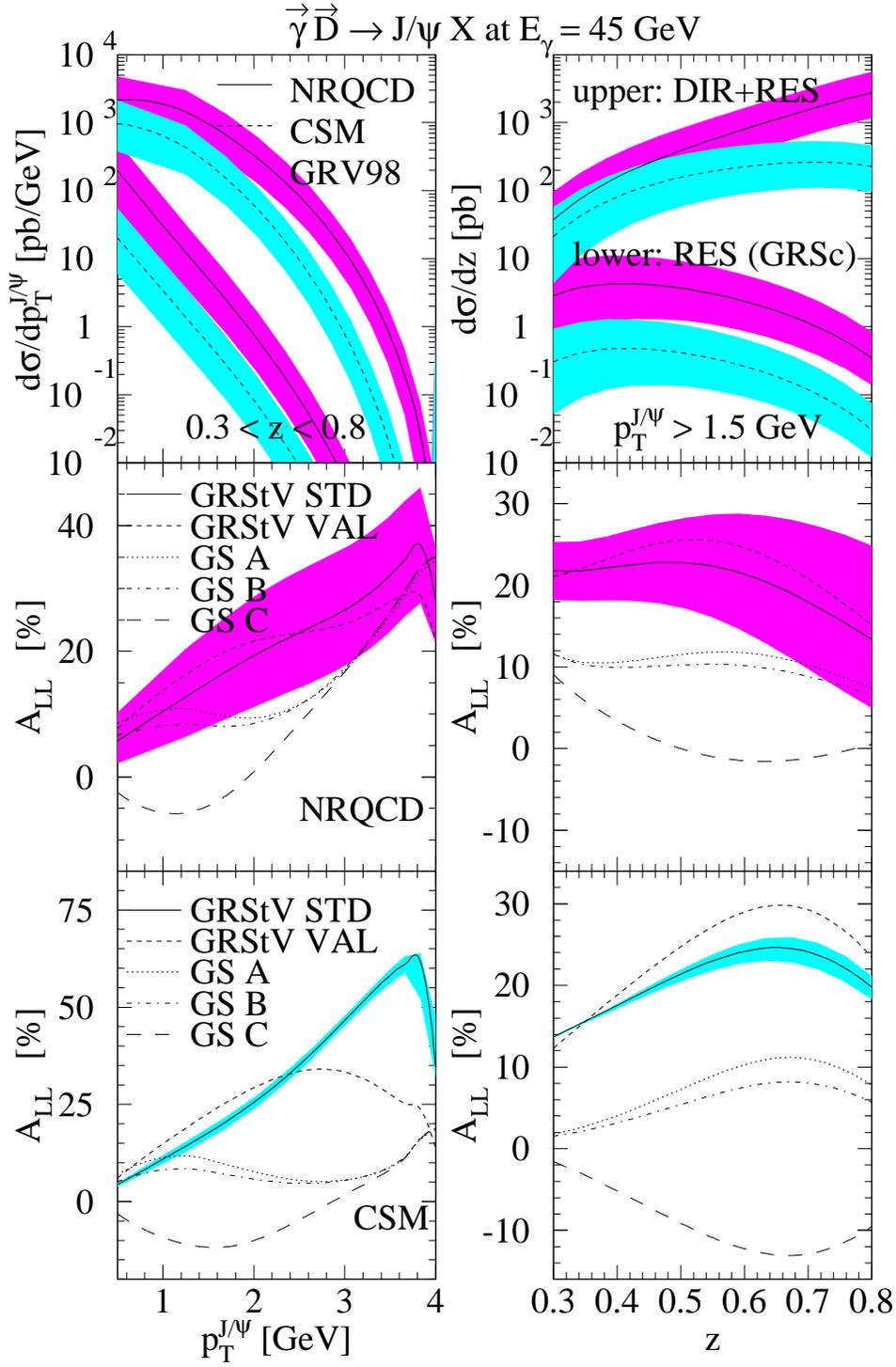,height=20cm}
\caption{Same as in Fig.~\ref{fig:rhic2}, but for $\gamma D\to J/\psi+X$ in
Experiment E161, with $E_\gamma=45$~GeV.
Inelasticity $z$ is used instead of rapidity $y$.
In the first panel, the resolved-photon contributions are also shown
separately.
\label{fig:slac}}
\end{center}
\end{figure}

\begin{figure}[ht]
\begin{center}
\epsfig{figure=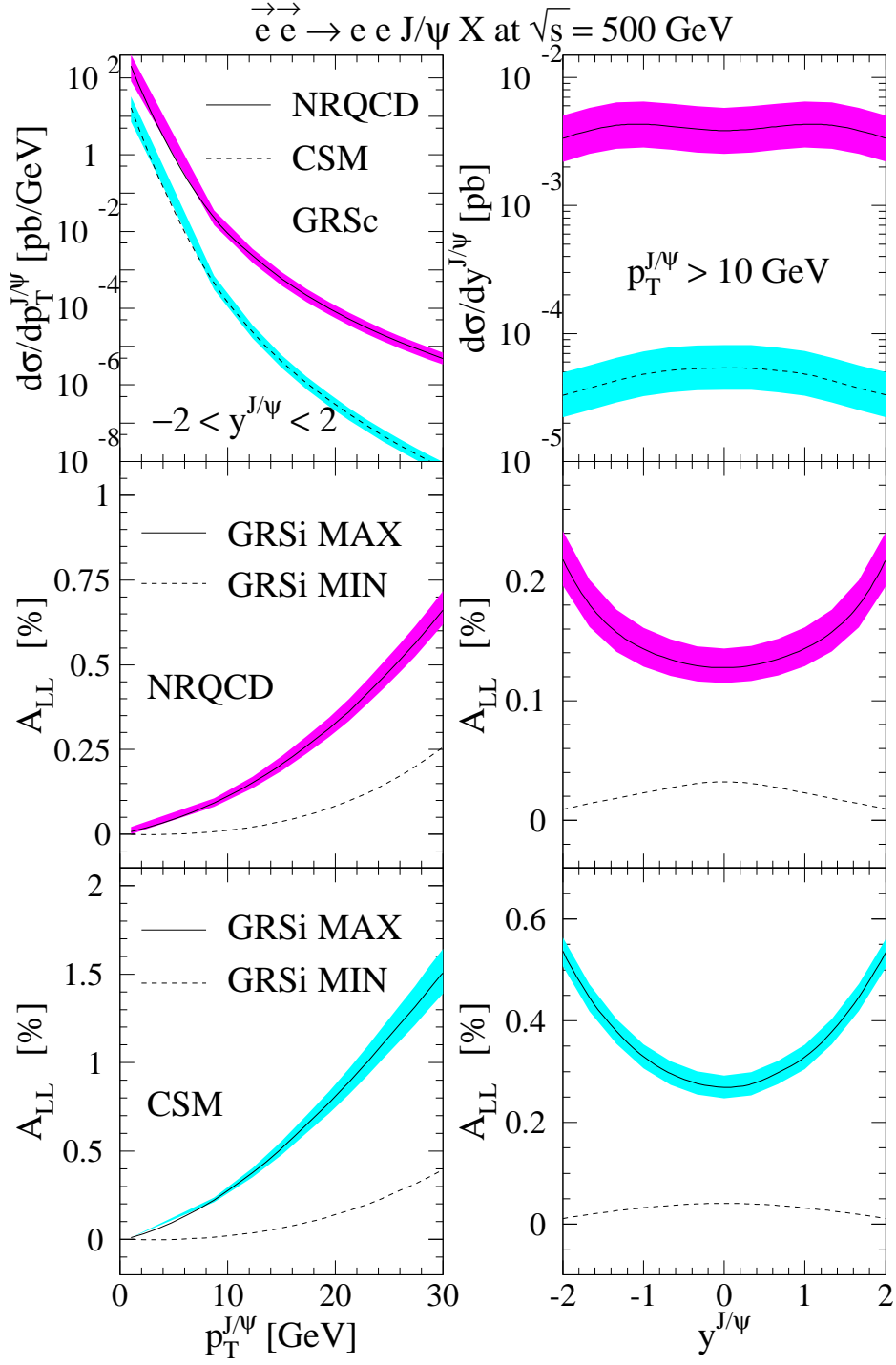,height=20cm}
\caption{Same as in Fig.~\ref{fig:rhic2}, but for $e^+e^-\to e^+e^-J/\psi+X$
in the $e^+e^-$ mode of TESLA, with $\sqrt S=500$~GeV, $P(e^+)=60\%$, and
$P(e^-)=80\%$.
\label{fig:tesla}}
\end{center}
\end{figure}

\begin{figure}[ht]
\begin{center}
\epsfig{figure=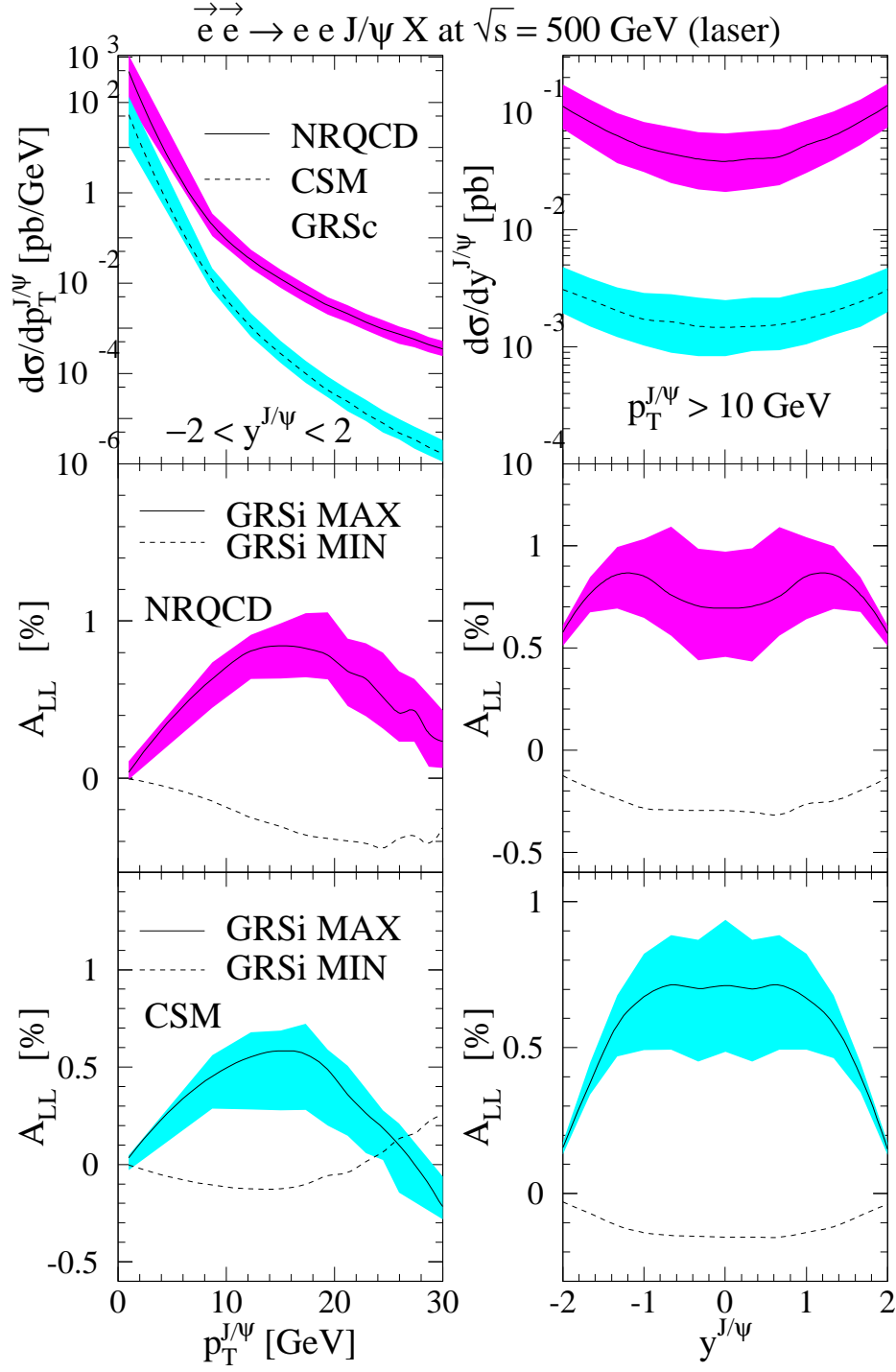,height=20cm}
\caption{Same as in Fig.~\ref{fig:rhic2}, but for $e^+e^-\to e^+e^-J/\psi+X$
in the $\gamma\gamma$ mode of TESLA, with $\sqrt S=500$~GeV, $P(e^-)=80\%$, 
and $P(\mbox{laser})=100\%$.
\label{fig:tesla.l}}
\end{center}
\end{figure}

Our main numerical results are presented in Figs.~\ref{fig:rhic2},
\ref{fig:slac}, \ref{fig:tesla}, and \ref{fig:tesla.l}, which refer to
$pp\to J/\psi+X$ at RHIC-Spin, to $\gamma D\to J/\psi+X$ in Experiment E161,
and to $e^+e^-\to e^+e^-J/\psi+X$ at TESLA in the $e^+e^-$ and $\gamma\gamma$
modes, respectively.
In each figure, the NRQCD and CSM predictions for the unpolarized cross
sections $d\sigma/dp_T$ and $d\sigma/dy$ or $d\sigma/dz$ are displayed in the
first panel, while those for the double longitudinal-spin asymmetry
${\cal A}_{LL}$ are shown in the second and third panels, respectively.
For the collider experiments, we consider $d\sigma/dy$, while for the
fixed-target scattering process $\gamma D\to J/\psi+X$, we consider
$d\sigma/dz$, where $z=(p_{J/\psi}\cdot p_D)/(p_\gamma\cdot p_D)
=\sqrt{m_{J/\psi}^2+p_T^2}\cosh y/E_\gamma$ is the inelasticity variable.
In Fig.~\ref{fig:slac}, which refers to the latter case, the first panel also
contains NRQCD and CSM predictions for the resolved-photon contributions by
themselves.
In the left and right columns of each figure, the $p_T$ and $y$ or $z$
dependences, respectively, are studied.
As for RHIC-Spin, E161, and TESLA, the $p_T$ distributions are integrated over
the intervals $|y|<2.4$, $0.3<z<0.8$, and $|y|<2$, while the $y$ or $z$
distributions are integrated over all kinematically allowed values of $p_T$ in
excess of 15, 1.5, and 10~GeV, respectively.
In Figs.~\ref{fig:rhic2}, \ref{fig:tesla}, and \ref{fig:tesla.l}, which refer
to collider experiments, the $y$ distributions are symmetric about $y=0$.
The shaded bands indicate the theoretical uncertainties in the NRQCD and CSM
default predictions, excluding the freedom in the choice of the PDFs.
In the case of ${\cal A}_{LL}$, these uncertainties are compared with the
spread due to a variation of the polarized PDFs.
We assume the ideal case of 100\% beam polarization, except for TESLA, where
we take $P(e^+)=60\%$ and $P(e^-)=80\%$.
As for RHIC-Spin and E161, realistic polarization is straightforwardly
accounted for by scaling ${\cal A}_{LL}$ with $[P(p)]^2$ and
$P(\gamma)P(D)$, respectively.

Considering the unpolarized cross sections, we observe in all cases that the
NRQCD predictions dramatically exceed the CSM ones, by typically one order of 
magnitude at small values of $p_T$.
This excess tends to amplify as $p_T$ increases, and it is striking in the $y$
distributions, thanks to the relatively large minimum-$p_T$ cuts chosen.
From the first panel of Fig.~\ref{fig:slac}, we observe that the 
resolved-photon contribution to $\gamma D\to J/\psi+X$ is greatly suppressed
in the entire kinematical range considered, both in NRQCD and the CSM.
For the same reason and because the polarized photon PDFs are poorly known at
present, we neglect this contribution in the evaluation of ${\cal A}_{LL}$
presented in the second and third panels of Fig.~\ref{fig:slac}.
By contrast, the NRQCD and CSM predictions for TESLA, shown in
Figs.~\ref{fig:tesla} and \ref{fig:tesla.l}, are greatly dominated by the
single-resolved contributions, as was already observed in Ref.~\cite{kkms}.
We conclude that, in all these experiments, the normalization of the
unpolarized cross section is a distinctive discriminator between NRQCD and the
CSM.

\begin{figure}[ht]
\begin{center}
\epsfig{figure=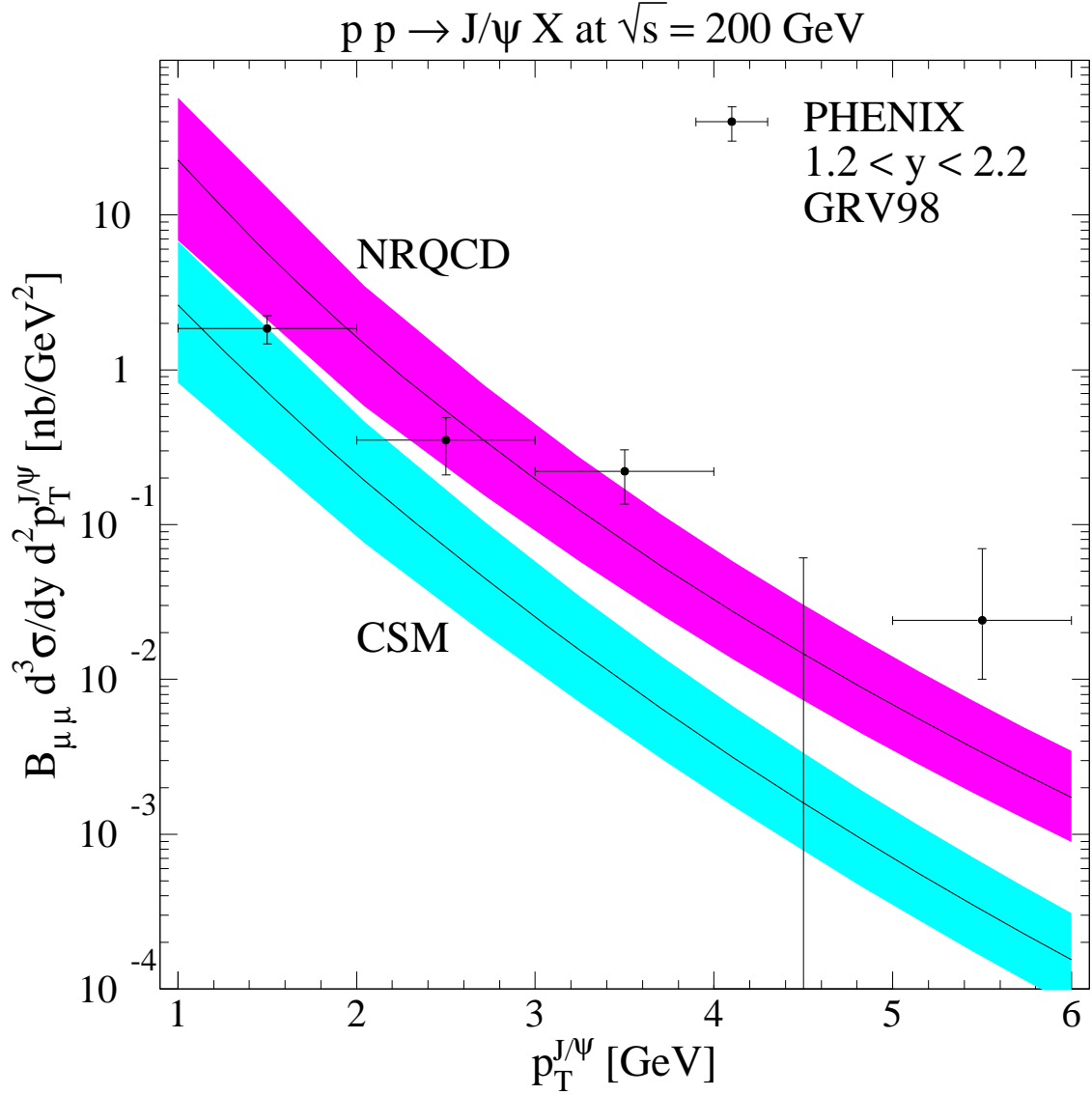,width=16cm}
\caption{The unpolarized cross section
$B(J/\psi\to\mu^+\mu^-)d^3\sigma/dy\,d^2p_T$ of $pp\to J/\psi+X$ followed by
$J/\psi\to\mu^+\mu^-$ measured by PHENIX \protect\cite{sat} at RHIC, with
$\sqrt S=200$~GeV, in the interval $1.2<y<2.2$ as a function of $p_T$ is
compared with the NRQCD and CSM predictions.
\label{fig:phenix}}
\end{center}
\end{figure}

In fact, data from the PHENIX Collaboration \cite{sat} at RHIC, with
$\sqrt S=200$~GeV, tend to favor the NRQCD prediction compared to the CSM one.
This is demonstrated in Fig.~\ref{fig:phenix}, where the differential cross
section $d^3\sigma/dy\,d^2p_T$ is analyzed for $1.2<y<2.2$ as a function of
$p_T$.
Since the $J/\psi$ mesons are tagged through their decays to $\mu^+\mu^-$ 
pairs, the factor $B(J/\psi\to\mu^+\mu^-)=(5.88\pm0.10)\%$ \cite{pdg} is
included in the theoretical predictions.
We observe that, for $p_T>2$~GeV, the data is nicely described by the NRQCD
prediction, while they significantly overshoot the CSM ones.
A similar observation was made in Ref.~\cite{nay}.
On the other hand, such a comparison has to be taken with a grain of salt in
the bin 1~GeV${}<p_T<{}$2~GeV because, at LO, the NRQCD prediction and the
$P$-wave contribution to the CSM one suffer from infrared and collinear
singularities at $p_T=0$, which still feed into that bin as an artificial
enhancement.

Looking at the second and third panels of Figs.~\ref{fig:rhic2},
\ref{fig:tesla}, and \ref{fig:tesla.l}, we observe that different choices of
polarized proton and photon PDFs yield discriminative NRQCD and CSM
predictions, as far as the colliding-beam experiments are concerned.
In fact, the differences are large against the combined theoretical
uncertainties from all other sources, especially at large values of $p_T$.
This means that sufficiently precise measurements of ${\cal A}_{LL}$ in these
experiments are bound to increase our knowledge on the spin-dependent parton
structure of the polarized proton and photon, regardless of the presently
still somewhat unsatisfactory status of the residual theoretical
uncertainties.
As for the $p_T$ distribution of ${\cal A}_{LL}$ at RHIC-Spin, the NRQCD and
CSM predictions incidentally almost coincide, so that the polarized proton
PDFs can be explored in a model-independent fashion.
On the other hand, the NRQCD and CSM predictions for the $y$ distribution of
${\cal A}_{LL}$ at RHIC-Spin exhibit strikingly different shapes in the
forward and backward directions.

As for the fixed-target experiment E161, the situation is somewhat less
favorable because the theoretical uncertainties not related to the polarized
proton PDFs are larger, owing to the relatively low photon-nucleon c.m.\
energy $\sqrt S=m_N(2E_\gamma+m_N)\approx9.2$~GeV.
This is especially the case for the NRQCD prediction shown in the second panel
of Fig.~\ref{fig:slac}.
Nevertheless, with enough experimental statistics, it should be possible to
discriminate between the GRStV and GS sets of polarized proton PDFs.
In the CSM, the resolving power of ${\cal A}_{LL}$ with respect to the
spin-dependent parton structure of the polarized proton used to be much 
better, as is evident from the third panel of Fig.~\ref{fig:slac}.
In this sense, the introduction of NRQCD, which was necessary to overcome
phenomenological and conceptual problems of the CSM, led to some aggravation.

\section{\label{sec:con}Summary}

Using the NRQCD factorization formalism at LO, we studied the inclusive
production of prompt $J/\psi$ mesons in polarized hadron-hadron,
photon-hadron, and photon-photon collisions at RHIC-Spin, experiment E161, and
TESLA, respectively, with regard to the potential of these facilities to
resolve the spin-dependent parton structure of the polarized proton and
photon.
For future use by other authors, we provided all contributing partonic cross
sections in analytic form.

Our main message is that the spread in the theoretical predictions encountered
by trying out in turn various up-to-date sets of polarized proton and photon
PDFs in general considerably exceeds the combined theoretical uncertainties
from other sources, which we estimated rather conservatively.
Therefore, these experiments have the power to improve our knowledge of the
spin structure of the proton and photon already at present.
This resolving power can be increased in the future by including 
next-to-leading-order corrections in $\alpha_s$ and $v$.

As a by-product of our analysis, we found that preliminary PHENIX data of
unpolarized hadroproduction of $J/\psi$ mesons at RHIC tends to favor NRQCD as
compared to the CSM.
This is in line with previous findings in connection with hadroproduction at
the Tevatron \cite{bra}, electroproduction at HERA \cite{dis}, and
photoproduction at LEP2 \cite{gg}.

\bigskip

\noindent
{\bf Acknowledgements}

\smallskip

\noindent
We thank Peter Bosted and Gerry Bunce for useful communications \cite{bos} and
Christoph Sieg for providing us with a parameterization of the GRSi polarized
photon PDFs \cite{grsi}.
This work was supported in part by the Deutsche Forschungsgemeinschaft through
Grants No.\ KL~1266/1-3 and KN~365/1-1, by the Bundesministerium f\"ur Bildung
und Forschung through Grant No.\ 05~HT1GUA/4, and by Sun Microsystems through
Academic Equipment Grant No.~EDUD-7832-000332-GER.

\renewcommand{\theequation}{\Alph{section}.\arabic{equation}}
\begin{appendix}
\setcounter{equation}{0} 

\section{Partonic cross sections} 

Here, we list the differential cross sections $d\sigma/dt$ of processes 
(\ref{eq:ppog})--(\ref{eq:qqog}) for arbitrary helicities $\xi_a,\xi_b$ of the
incoming particles.
The Mandelstam variables $s$, $t$, and $u$ are defined in the usual way, and
$M=2m_c$, where $m_c$ is the charm-quark mass.
By four-momentum conservation, we have $s+t+u=M^2$.
Furthermore, $e=\sqrt{4\pi\alpha}$ and $g_s=\sqrt{4\pi\alpha_s}$ are the
electromagnetic and strong gauge couplings, and $Q_q$ is the fractional
electric charge of quark $q$.
Our results read:



\end{appendix}

\newpage

\newpage

\newpage

\newpage

\newpage

\end{document}